\documentclass[twocolumn]{revtex4}
\usepackage{graphicx}
\usepackage{amssymb}

\begin{document}

\preprint{{\bf MADPH-04-1368}}

\title{Direct experimental test of scalar confinement}
\author{Theodore J. Allen}
\affiliation{Department of Physics, Hobart \&
William Smith Colleges, \\ Geneva, New York 14456, USA}

\author{M.G. Olsson and Yu Yuan} \affiliation{Department of Physics,
University of Wisconsin, 1150 University Avenue, Madison, Wisconsin 53706, USA}

\author{Jeffrey R. Schmidt} 
\affiliation{Department of Physics, University of Wisconsin-Parkside,\\ 
900 Wood Road, Kenosha, Wisconsin 53141, USA}

\author{Sini\v{s}a Veseli} 
\affiliation{Fermi National Accelerator
Laboratory, P.O. Box 500, Batavia, Illinois 60510, USA}

\begin{abstract} 
The concept of Lorentz scalar quark confinement has a long history and is
still widely used despite its well-known theoretical faults.  We point out
here that the predictions of scalar confinement also conflict directly with
experiment.  We investigate the dependence of heavy-light meson mass
differences on the mass of the light quark.  In particular, we examine the
strange and non-strange $D$ mesons.  We find that the predictions of scalar
confinement are in considerable conflict with measured values.

\end{abstract}

\maketitle

\section{A little history}
Within a year of the discovery of charmonium in 1974, a picture emerged of
two massive quarks moving in the QCD motivated static potential
\cite{Cornell},
\begin{equation}
                   V(r)=-\,\frac{k}{r}+ar .
\label{Cornell}
\end{equation}
Here the color Coulomb constant $k$ is given by $k=4\alpha_s/3$ in lowest
order perturbation theory and $a$ is the confinement constant, or
asymptotic confining force.  The non-relativistic Schr\"odinger equation
accounted in a natural way for the spin averaged $c\bar c$ and $b\bar b$
spectra and decay rates \cite{Buchmuller}.  It was soon realized that in
order to understand the spin dependence of the heavy onia states one must
further specify the Lorentz transformation properties of the interaction.
Henriques, Kellett, and Moorhouse \cite{Henriques} proposed that the short
range Coulombic part was a Lorentz vector and the long range confining part
a Lorentz scalar.  The reasoning by the above group and others
\cite{Buchmuller,Nathan} was that in order to account for observations, the
spin-orbit interaction must be suppressed by a partial cancelation between
the short-range and long-range contributions.

This picture flourished unchallenged for fifteen years despite the lack of
success in formally relating scalar confinement to QCD.  Around 1990, the
work of the Milan group \cite{Milan} clarified this question dramatically.
Using the low velocity Wilson loop formalism pioneered by Eichten and
Feinberg \cite{E&F}, and later by Gromes \cite{Gromes}, they found both the
spin-dependent and spin-independent relativistic corrections in heavy onia.
Their astounding result was that the long-range spin-independent QCD
corrections differed from those of scalar confinement, though the
long-range spin-orbit corrections were the same pure Thomas ones given by
scalar confinement.  These results were subsequently verified by lattice
simulations of QCD \cite{Bali}.

Because the spin-independent relativistic corrections to scalar confinement
are incorrect, the scalar confinement scenario should logically be
discarded. Indeed, the whole concept of potential confinement is in error.
Fortunately, there is an alternative physical picture that can be employed.
Back in 1982, Buchm\"uller \cite{Buch} pointed out that a color electric flux
tube should automatically yield the desired pure Thomas spin-orbit
interaction because there is no color magnetic field in the quark rest
frame.  In 1992, it was shown that the spin-independent relativistic
corrections \cite{Ken} to the flux tube model exactly matched the QCD
predictions \cite{Milan}.  Recently, we have constructed a consistent
classical action for spinning quarks connected by a QCD string (flux tube)
\cite{spin}. Hence there exists a simple physical picture that is
consistent in many ways with QCD and is not a simple potential model.

Although inconsistent with QCD, scalar confinement models remain popular,
probably because of their relative ease of solution.  In this paper we
point out an instance where the predictions of spin-independent scalar
confinement disagree directly with experiment.  We consider here the
dependence of heavy-light (HL) meson masses on the light quark mass.  We
discuss a general HL potential model wave equation in
Sec.~\ref{sec:perturb} and exhibit a simple analytic perturbative solution
for scalar confinement. We demonstrate the high accuracy of our numerical
solutions by comparing them to the analytic ones.  In
Sec.~\ref{sec:experiment} we collect and discuss the experimental data
which we will compare to our predictions.  In particular, we will establish
that spin splittings do not depend on the mass of the light quark.
Comparison of three confinement scenarios to data is given in
Sec.~\ref{sec:comparison}, where we give theoretical values obtained by
exact numerical solution of the spinless Salpeter equation for both scalar
and time component vector (TCV) confinement as well as flux tube
confinement.  Our conclusions in Sec.~\ref{sec:discussion} are that scalar
confinement predictions of the light quark mass dependence are not good and
can be clearly seen in the difference of \textit{S}-wave and
\textit{P}-wave meson states.  We note further that electric (TCV)
confinement also does not account for this difference but that the flux
tube model exhibits remarkable agreement with experimental data.

\section{Scalar confinement equation:  a perturbative solution and an exact
numerical solution}\label{sec:perturb}

\subsection{The heavy-light potential model}\label{subsec:HL}

We consider a spinless quark of mass $m$ that moves in a potential field.
We assume that this field consists of central Lorentz scalar and time
component vector fields, $S(r)$ and $V(r)$ respectively.  The expected
Lagrangian for this system is
\begin{eqnarray}
            L & = & -m(r)\sqrt{1-\mathbf{v}^2} - V(r) ,    \label{lag} \\
            m(r) &= & m+S(r) .                              \label{m(r)}
\end{eqnarray}
The momentum and Hamiltonian are then
\begin{eqnarray}
          \mathbf{p} & = & m(r) \, \gamma \, \mathbf{v} ,      \label{p} \\
                   H & = & \mathbf{v}\cdot\mathbf{p} -  L  ,   \label{Ham1}
\end {eqnarray} 
where $\gamma  =  1/\sqrt{1-{\bf v}^2}$.

By squaring $\bf{ p}$, we see that
\begin{equation}
        m(r) \, \gamma = \sqrt{\mathbf{p}^2+m(r)^2} .         \label{gam1}
\end{equation}
Substitution into the Hamiltonian (\ref{Ham1}) then yields
\begin{equation}
              H= \sqrt{\mathbf{p}^2+(m+S(r))^2}+V(r)   .      \label{Ham2}
\end{equation}

For a state of definite angular momentum ${\ell}$ the momentum square can
be written
\begin{equation}
          {\bf p}^2={p_r}^2+{\ell(\ell +1)\over r^2}  .          \label{psph}
\end{equation}

With the usual replacement $p_r^2 \to -\,\frac{1}{r}\frac{d^2  }{d r^2} r$, we
obtain a wave equation that can always be solved numerically \cite{Galerkin}.

\subsection{A perturbative estimate of the light quark mass dependence}\label{subsec:pert}

We next consider a simple model with scalar confinement,
$S(r)=a|\mathbf{r}|=a r$ and a time component vector short range
interaction, $V(\mathbf{r})=-{k\over |\mathbf{r}|}=-{k\over r}$.
From the general Hamiltonian (\ref{Ham2}),  we find

\begin{equation}
        H= \sqrt{{\bf p}^2+(m+ar)^2} - \frac{k}{r} .        \label{HamS}
\end{equation}
We wish to treat the short range parameter $k$ and the light quark mass $m$
as small perturbations.  In the case where both $k$ and $m$ vanish, the
zeroth order Hamiltonian is
\begin{equation}
                           H_0= \sqrt{{\bf p}^2+(ar)^2} .        \label{HamS0}
\end{equation}

The Hamiltonian $H_0$ has the same eigenstates as its square which, with the
replacements of Eq.~(\ref{psph}) and $p_r^2 \to -\,\frac{1}{r}\frac{d^2}{d
r^2 }r$, leads to the harmonic oscillator equation,
\begin{equation}
\frac{d^2u_0}{d r^2} + 
\left(E_0^2 - {\ell(\ell +1)\over r^2} - a^2r^2\right)u_0 = 0 ,
\label{HO}                         
\end{equation}
where in the limit of small $r$, $u_0\to r^{\ell+1}$  and $u_0$ is
normalized to
\begin{equation}
\int_0^{\infty} dr \, {\left| u_0 \right| }^2  =1 .    \label{intnorm}
\end{equation}
The solution for the wave function and eigenvalue is standard;
\begin{eqnarray}
u_0(r) &=& \mathcal{N}_{n,\ell}\, r^{\ell+1}\, e^{-\frac{1}{2}ar^2}\, 
L_{n-1}^{\ell+\frac{1}{2}}(ar^2)  ,      \label{soln0}\\
 \mathcal{N}_{n,\ell}^2 &=& \frac{2a^{\ell +\frac{3}{2}}\, (n-1)!}
{\Gamma(\ell+n+\frac{1}{2})} ,
\label{norm}\\
%% See p 84, N. N. Lebedev, ``Special functions''
E_0^2 & = & 2a\,\left(\ell+2n-\frac{1}{2}\right) .          \label{eigen0}
\end{eqnarray}
Here $L_{n-1}^{\ell+\frac{1}{2}}(ar^2) $ is the usual Laguerre
polynomial, $n$ is a positive integer starting with $1$, and $\ell$ is a
non-negative integer starting with zero.

We now determine the effect of turning on $k$ and $m$ using the
Feynman-Hellmann theorem \cite{FH},
\begin{equation}
\partial E/\partial \lambda = \langle \partial H/\partial \lambda \rangle  ,
\label{FH}
\end{equation}
where $\lambda$ is any parameter of the Hamiltonian.  Taking the
expectation values using the $k=m=0$ wavefunctions (\ref{soln0}) will yield
an expansion in these parameters.  To leading order we have
\begin{equation}
 E=E_0-k \, \left\langle {r^{-1}}\right\rangle +\frac{ma}{E_0}\,
\langle r \rangle +\dots  \ .    \label{expan}
\end{equation}

The expectation values are worked out in general in the Appendix.  Here we
only consider the $m$ and $k$ dependence of the 1\textit{S} and 1\textit{P}
states (i.e., $n=1$ and $\ell=0$ and $1$).  The results are,
\begin{eqnarray}
\langle {r^{-1}} \rangle\strut_{1S} 
& = &2 \sqrt{\frac{a}{\pi} } , \label{1S/r}\\
\langle {r^{-1}} \rangle\strut_{1P} 
& = &\frac{4}{3} \sqrt{\frac{a}{\pi} }, \label{1P/r}\\
\langle r \rangle\strut_{1S} & = & \frac{2}{\sqrt{\pi a}}  , \label{1Sr}\\
\langle r \rangle\strut_{1P} & = & \frac{8}{3\sqrt{\pi a}} . \label{1Pr}
\end{eqnarray}

Using Eq.~(\ref{expan}), we obtain the $m$ dependence near $k=m=0$ ($m$
slope) for the energies,
\begin{eqnarray}
\frac{d\,E_{1S}}{dm} & = &  \frac{2}{\sqrt{3\pi}} =0.6515 ,  \label{mE1S} \\
\frac{d( E_{1P}-E_{1S})}{dm} 
& = &\frac{2} {\sqrt{3\pi}}\left(\frac{4}{\sqrt15}-1\right)= 0.0214 .
\label{mdiff}
\end{eqnarray}  
Again using Eq.~(\ref{expan}) we obtain the $k$ dependence near $k=m=0$
($k$ slope) for the energies,
\begin{eqnarray}
 \frac{1}{\sqrt a } \frac{d\,E_{1S}}{dk} 
& = & -2\sqrt{\frac{1}{\pi} } =~-1.128 ,  \label{kE1S} \\
 \frac{1}{\sqrt a } \frac{d( E_{1P}-E_{1S})}{dk} 
& = &\frac{2}{3}\sqrt{\frac{1}{\pi} } = 0.376 .  \label{kdiff}
\end{eqnarray}

\subsection{Consistency of the exact numerical solution and the analytic
perturbative solution} \label{subsec:consistency}

Before we proceed to a detailed comparison of the predictions of scalar
confinement with the experimental data, we pause to verify the accuracy of
our numerical method.  In particular, we compare our analytic values for
the slopes at $m=k=0$ in Eqs.~(\ref{mE1S}) to (\ref{kdiff}) with our
numerical method.  This step is important because a general analytic
solution of the spinless Salpeter equation is not known.  Only for the
remarkable case of a massless particle and linear scalar confinement, which
is equivalent to the non-relativistic harmonic oscillator, can one obtain
an analytic solution.  In the general case, one must rely on exact
numerical solutions.  Some time ago, we introduced \cite{Galerkin} a
variational method, the Galerkin method, into particle physics to solve the
spinless Salpeter equation with a time component vector interaction.  This
very robust method is applicable to a wide range of differential and
integral equations.  The method has been sharpened over the years by many
authors \cite{others}.  One can cope with eigenvalue equations for
operators that are complicated functions of both momenta and coordinates,
such as the scalar confinement Hamiltonian (\ref{HamS}), by using basis
functions that can be Fourier transformed.  We have performed a careful
numerical solution of the eigenvalue equation for the Hamiltonian
(\ref{HamS}), for small $k$ and $m$, and found the $m$ and $k$ slopes. The
results are in excellent agreement with the values obtained by analytic
calculation in Eqs.~(\ref{mE1S}) through (\ref{kdiff}).

\section{Experimental data} \label{sec:experiment}

\subsection{Spin splitting is independent of light quark mass}
\label{subsec:spinsplit} 

We first use experimental data to demonstrate, rather conclusively, that
spin splittings within a given orbital angular momentum multiplet do not
depend on the light quark mass.  In particular, we consider several
heavy-light meson spin multiplets in which both the strange and non-strange
members have been observed \cite{PDG}.  First we examine the $D$ and $D_s$
type mesons with $\ell =0$ (\textit{S}-waves).  The hyperfine splittings
for $D_s$ and $D$ mesons are
\begin{eqnarray}
        D_s^* - D_s & = & 143.8~\pm 0.4~\textrm{MeV},  \nonumber     \\
        D_\pm^* - D_\pm & = & 140.64~\pm~0.10~\textrm{MeV}, \label{Dhyp} \\
        D_0^* - D_0 & = &  142.12~\pm 0.07~\textrm{MeV}. \nonumber
\end{eqnarray}
The corresponding \textit{S}-wave hyperfine splittings for $B$ type states are
\begin{eqnarray}
        B_s^* - B_s & = & 47.0~\pm 2.6 ~\textrm{MeV}, \nonumber \\
           B^* - B  & = &  45.78~\pm 0.35~\textrm{MeV}.        \label{Bhyp}
\end{eqnarray}
It is clear that the substitution of a strange quark for a non-strange
light quark changes the \textit{S}-wave hyperfine differences by at most a few MeV.

We next consider some measured \textit{P}-wave heavy-light spin splittings.  From
\cite{PDG} we find,
\begin{eqnarray}
      D_{s2} - D_{s1}    & = & 37.0~\pm 1.6~\textrm{MeV} , \nonumber \\
      D_2^0 - D_1^0      & = & 36.7~\pm 2.7~\textrm{MeV} ,   \label{DP} \\
      D_2^\pm  - D_1^\pm & = & 32~\pm 6 ~\textrm{MeV} . \nonumber
\end{eqnarray}

Again we note the apparent vanishing of light quark mass dependence this
time in a \textit{P}-wave spin splitting.  We conclude that spin splittings are only
weakly dependent on the light quark mass.  We will exploit this fact in
section \ref{subsec:conclusion}.

\subsection{The light quark mass dependence of a 1\textit{P}$-$1\textit{S} difference} 
\label{subsec:massdiff}

In the preceding subsection we observed from experiment that both \textit{S}-wave and
\textit{P}-wave heavy-light spin splittings (within a spin multiplet) were
independent of the light quark mass.  We next consider the mass splittings
between pairs of states
corresponding to different orbital angular momenta and examine the light
quark mass dependence of this difference.  We choose the $D_1$ \textit{P}-wave state
and the pseudoscalar $D$ meson.  The best measurements are
\begin{equation}
  \Delta_u \ = \   D_1^0 - D^0 \ = \  557.5~\pm 2 ~ \textrm{MeV}.\label{SPD}
\end{equation}
When  the $u$ light quark is replaced by a strange quark, the corresponding
difference becomes
\begin{equation}
  \Delta_s \ = \ D_{s1} - D_s \ = \ 567.3~\pm 0.4~ \textrm{MeV}.\label{SPSD}
\end{equation}
The differences $\Delta_u$ and $\Delta_s$ are amazingly similar.  We see that
they differ by
\begin{equation}
 \Delta \ = \ \Delta_s -  \Delta_u \ = \ 9.8~\pm 2~\textrm{MeV}. \label{Diff}
\end{equation}

\subsection{Conclusion for the spin-averaged 1\textit{P}$-$1\textit{S} difference}
\label{subsec:conclusion}

We demonstrated in Sec.~\ref{subsec:spinsplit} that both \textit{S}- and
\textit{P}-wave spin splittings are, within error and isospin uncertainty,
independent of light quark mass.  We may therefore conclude that the
$m$-dependence of the difference $ \Delta_s- \Delta_u$ also represents the
$m$-dependence of the spin-averaged 1\textit{P}$-$1\textit{S} excitation
energies.  To show this explicitly, we write the HL meson mass as the sum
of the heavy quark mass and the excitation energy,
\begin{equation}
M=m_Q +E .
\end{equation}
We then separate the excitation energy into spin-averaged and
spin-dependent parts,
\begin{equation}
E=E^{SA} +E^{SD} .
\end{equation}
As in Eqs.~(\ref{SPD}) and (\ref{SPSD}), we define $\Delta_s$ and
$\Delta_u$ to be the differences between \textit{P}-wave ($J^P = 1^+$) and
\textit{S}-wave ($J^P = 0^-$) states, for strange and non-strange light quarks
respectively.  Using our observation that the spin-dependent parts are
essentially independent of light quark mass, we see that the spin dependent
parts will cancel in the difference $\Delta ={\Delta_s} -{\Delta_u}$. Thus
\begin{equation}
\Delta = (E_{1P} -E_{1S})_s -  (E_{1P} -E_{1S})_u 
\end{equation}
measures the light quark mass dependence of the spin-independent 1\textit{P}$-$1\textit{S}
level difference.

Two main results emerge from this subsection.  First, the light quark mass
dependence is only about 2\% of the measured level difference. This is a
useful tool in the analysis of heavy-light spectroscopy \cite{universal}.
Second, although it is small, $\Delta$ is definitely not zero.  We will use
the actual value (\ref{Diff}) to test the predictions of various
assumptions about the nature of quark confinement.

\section{Comparison with experiment} 
\label{sec:comparison}

\begin{figure}[h]
\includegraphics[width=\columnwidth,angle=0]{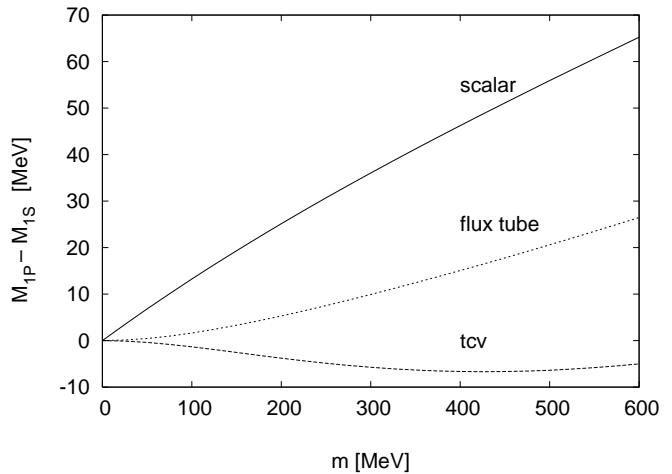}
\caption{Dependence of the 1\textit{P}$-$1\textit{S} energy level difference on the light quark
mass $m$. The numerical results compare different confinement mechanisms
(all with the same long range confinement force $a=0.18$~GeV$^2$ and all
three calculations assume the same short range Coulombic constant
$k=0.5$.)\label{fig:one}}
\end{figure}

We are now prepared to compare carefully the different confinement
predictions with experiment.  As we originally noted \cite{universal}, the
$m$ dependences of all the heavy-light states are amazingly similar.  We
have also noted in Sec.~\ref{subsec:consistency} that this universal $m$
dependence is nearly satisfied in an analytical calculation.  In the
example with scalar confinement, the $m$ slope for the difference 1\textit{P}$-$1\textit{S} is
about 30 times smaller than each separate slope.  Furthermore, we noted in
Sec.~\ref{subsec:massdiff} that when one compares two heavy-light states, one
\textit{P}-wave and one \textit{S}-wave, the difference changes by less
than 2\% when a
non-strange light quark is replaced by a strange one.  This
change is not zero however, but for the $D_1$ and $D$ states has the
experimental value (\ref{Diff})
\begin{equation}
  \Delta \ = \ \Delta_s - \Delta_u \ = \ 9.8~\pm 2~\textrm{MeV} .
\end{equation}
In Fig.~\ref{fig:one} we show the numerical mass splittings of heavy-light
mesons for three different confinement scenarios, all with the same short
range energy $-k/r$, and the same $k=0.5$, as a function of light quark
mass. Each of the three confinement scenarios has the same asymptotic
confinement force, $a=0.18$~GeV$^2$.  The upper curve assumes linear scalar
confinement, the middle curve is the prediction of the relativistic flux
tube and the lower curve linear time component vector confinement. In the
scalar and  time component vector potentials, the potentials $S(r)$
and the long-range part of $V(r)$ respectively are $ar$.  In the case
relativistic flux tube model the string tension is $a$.
 
\begin{figure}[h]
\includegraphics[width=0.14\columnwidth]{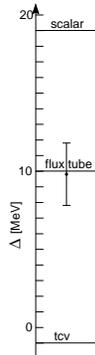}
\caption{The difference in the 1\textit{P}$-$1\textit{S} energy gap between a light quark
mass value of 500~MeV and 300~MeV. The experimental value is the dot below
the 10~MeV level. All theoretical predictions have the same long range
confinement force, $a=0.18$~GeV$^2$, and the same short range Coulombic
constant, $k=0.5$.\label{fig:two}}
\end{figure}

We note that the scalar confinement has the most rapid increase of the
1\textit{P}$-$1\textit{S} difference as the light quark mass increases. The
flux tube confinement is intermediate and time component vector linear
confinement actually decreases slightly.  Using the reasonable values for
the strange quark mass (500 MeV) and the non-strange mass (300 MeV)
\cite{universal}, we find the following values for $\Delta$,
\begin{eqnarray}
     \Delta_{\rm scalar}\ =\ 19~\textrm{MeV} ,   \\
     \Delta_{\rm flux~tube}\ =\ 10~\textrm{MeV} ,\\
     \Delta_{\rm TCV}\ =\ -1~\textrm{MeV}  .    
\end{eqnarray}
In comparing these values  to the experimental value given in Eq.~(\ref{Diff}),
\begin{equation}
     \Delta_{\rm exp}=~9.8~\pm 2~\textrm{MeV} ,
\end{equation}
we observe that both scalar confinement and time component vector
confinement are quite inconsistent with the experimental result.  However,
the flux tube confinement prediction is in agreement with the experimental
result.  These results are depicted in Fig.~\ref{fig:two}.

\section{Discussion}    \label{sec:discussion}

\subsection{$m$ slopes and the $k$ value}

The reader may have noticed that the $m$ slope for the
1\textit{P}$-$1\textit{S} difference calculated analytically (\ref{mdiff})
is about 0.02 whereas the $m$ slope in Fig.~\ref{fig:one} is about five
times larger.  This is because the Coulomb constant $k$ is zero in the
analytical result and has the more realistic value 0.5 in the numerical
calculations shown in Fig.~\ref{fig:one}. The effect is magnified due to
the strong cancelation in the 1\textit{P}$-$1\textit{S} difference.

\begin{figure}[h]
\includegraphics[width=\columnwidth,angle=0]{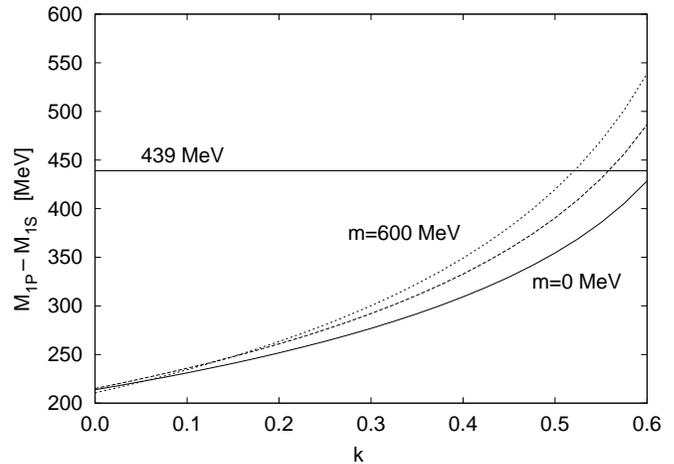}
\caption{The 1\textit{P}$-$1\textit{S} level difference as a function of
the Coulomb constant $k$ for scalar confinement.  In this case there is a
noticeable dependence on the light quark mass $m$ and we show the
calculation for $m=0$, $300$, and $600$ MeV.  Since the experimental value
was found for $D_s$ mesons we see a value of $k=0.53$ is indicated with
$m=500$ MeV.\label{fig:three}}
\end{figure}

The choice $k=0.5$ is quite reasonable.  As we noted in \cite{universal}, the
spin-averaged experimental value for $E_{1P}-E_{1S}$ can be extracted from
the $D_s$ mesons;
\begin{equation}
E_{1P} - E_{1S}\ =\ 439\ \textrm{MeV}.
\end{equation}
The appropriate choice of $k$ is determined by computing the above quantity
as a function of $k$ as shown in Fig.~\ref{fig:three}.
We conclude that the common value of $k=0.5$ assumed in the considerations of
the last section was appropriate and that the effect of a slightly larger $k$
would only make our conclusions stronger.

\subsection{Constituent quark masses}

Another interesting result of a choice of confinement scenario is a
constraint that relates the non-strange and strange light quark masses.
This relation is the result of the measured value for the $B_s$ and $B$
mass difference,
\begin{equation}
           B_s - B\ \simeq\ 91\ \textrm{MeV}.       \label{Bdiff}
\end{equation}

\begin{figure}[h]
\includegraphics[width=\columnwidth,angle=0]{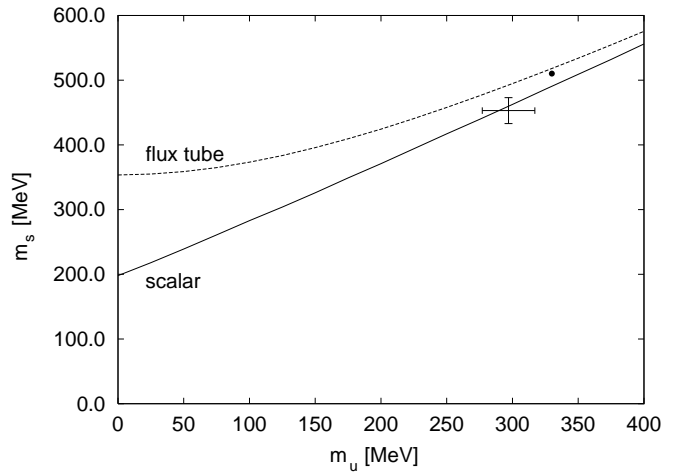}
\caption{The constraint between $m_s$ and $m_u$ light quark masses due to
the measured $B_s-B$ mass difference (\ref{Bdiff}).  The solid curve
assumes scalar confinement while the dashed curve is due to flux-tube
confinement.  The ``data'' points are quark masses analyses of hyperon
magnetic moments in the constituent quark model
\cite{PDG,hypmag}. \label{fig:four}}
\end{figure}

In Fig.~\ref{fig:four} we show this constraint scalar confinement and flux
tube confinement scenarios.  Although the curves differ considerably for
small light quark mass, they are quite similar in the larger range.  On the
figure we show constituent quark masses obtained by analyses of hyperon
magnetic moments \cite{PDG,hypmag}.  The result justifies our choice of 500
MeV for the strange quark and 300 MeV for the non-strange quark.

\section{Conclusions}

Our points in this paper may be quickly summarized.

\begin{itemize}

\item From the data, we conclude that spin splittings do not vary with
light quark mass value.  We then can extract the change (\ref{Diff}) in the
spin-averaged 1\textit{P}$-$1\textit{S} level difference as one replaces
the non-strange by a strange light quark.

\item We compare the predictions of the scalar potential, time
component vector potential and flux-tube quark confinement scenarios with
experimental results, as shown in Fig.~\ref{fig:two}.  The conclusion is
that flux-tube confinement works well while both scalar and time component
vector confinement fail badly.

\end{itemize}

We observe therefore that scalar confinement has at least one specific
point of disagreement with experiment.  This complements the theoretical
disagreements with QCD mentioned in the introduction.

\section*{Acknowledgment}
This work was supported in part by the U.S. Department of Energy under
Contract No.~DE-FG02-95ER40896.

\appendix

\section{Expectation Values}

The principal aim here is to compute the expectation value of $r^p$ with
the harmonic oscillator wavefunctions (\ref{soln0}),
\begin{equation}
\langle r^p \rangle =N_{n,\ell}^2\> 
\int_0^{\infty} dr\,r^{2\ell+2+p}\, e^{-ar^2}\,
\left[L_{n-1}^{\ell+\frac{1}{2}}(ar^2)\right]^2 .   \label{exr}
\end{equation}
A change of integration variable to the dimensionless combination
$z=a\,r^2$ yields 
\begin{equation}
2\,a^{\ell +\frac{3}{2}+\frac{p}{2}}\, \langle r^p \rangle = N_{n,\ell}^2\,
\int_0^\infty dz\,z^{\ell
+\frac{1}{2}+\frac{p}{2}}\,e^{-z}\,\left[L_{n-1}^{\ell
+\frac{1}{2}}(z)\right]^2 .        \label{exz}
\end{equation}
It is helpful to use the Chu-Vandermonde sum formula \cite{GR},
\begin{equation}
L_{n-1}^\alpha (z)=\sum_{j=1}^n ~\frac{(\alpha -\beta)_{n-j}}{(n-j)!}
~L_{j-1}^\beta (z) ,
\label{CV}
\end{equation}
where the Pochhammer symbol $(z)_N$ is defined as,
\begin{eqnarray}
(z)_N =z(z+1)\cdots (z+N-1) & = & \frac{\Gamma(z+N)}{\Gamma(z)},\label{poch}\\
(z)_0 & = & 1 .
\end{eqnarray}

With the choices
\begin{eqnarray}
             \alpha & = & \ell +\frac{1}{2} , \\  
              \beta & = & \ell +\frac{1}{2} + \frac{p}{2} ,    \\  \label{def1}
      \alpha -\beta & = & - \,\frac{p}{2} ,
\end{eqnarray}
we substitute Eq.~(\ref{CV}) into (\ref{exz}) and use the orthonormality
relation for Laguerre polynomials \cite{GR}
\begin{equation}
\int_0^\infty dz\,z^\beta \, e^{-z} \, L_j^\beta (z) \, L_{j'}^\beta (z) =
\frac{\Gamma(j+\beta + 1)}{j!} \, \delta_{jj'}   ,   \label{orthog} 
\end{equation} 
to obtain our general result,
\begin{equation}
\langle r^p \rangle = a^{-\frac{p}{2}} \, \sum_{j=1}^n
\,\left[\left(-\frac{p}{2}\right)_{n-j}\right]^2\, \pmatrix{n-1 \cr j-1}
\frac{\Gamma (j+\ell+\frac{1}{2}+\frac{p}{2})}{\Gamma
(n+\ell+\frac{1}{2})(n-j)!} .
\label{finalexpectation}
\end{equation}
In Eq.~(\ref{finalexpectation}) we use the notation for the binomial coefficients,
\begin{equation}
\pmatrix{n\cr m} = \frac{n!}{m!(n-m)!} .      \label{binomial}
\end{equation}
The specific result that is required in Sec.~\ref{subsec:pert} is for the
ground state 
($n=1$) is
\begin{equation}\label{n=1}
\langle r^p \rangle\strut_{n=1} = a^{-\frac{p}{2}} ~\frac{\Gamma (\ell
+\frac{3}{2}+\frac{p}{2}) }{\Gamma(\ell+\frac{3}{2})} 
\end{equation} 
Some cases of direct interest are
\begin{eqnarray}
  & p = -1: \qquad &\left\langle {r^{-1}} \right\rangle = \sqrt{a}\,
              \frac{\Gamma(\ell+1)}{\Gamma(\ell+\frac{3}{2})} , \\ 
  & p =  0: \qquad &\langle 1 \rangle = 1 , \\
  & p =  1: \qquad &\langle r \rangle = \frac{1}{\sqrt{a}}\,
              \frac{\Gamma(\ell+2)}{\Gamma(\ell+\frac{3}{2})} , \\ 
  & p =  2: \qquad &\langle r^2 \rangle = \frac1a \left(\ell +\frac{3}{2}\right) .
\end{eqnarray}

\end{document}